\documentclass[10pt]{article}
\usepackage{graphicx}
\usepackage{amsmath}
\makeatletter

\begin{document}

\title{\textbf{ Comment on "Classification scheme for phenomenological universalities in growth problems in physics and other sciences"}}

\author{\textbf{Dibyendu Biswas$^1$}\footnote{dbbesu@gmail.com} \textbf{and Swarup Poria$^2$}\footnote{swarupporia@gmail.com}\\
\small{$^1$Department of Basic Science, Humanities and Social Science (Physics)}\\\small{ Calcutta Institute of Engineering and Management }\\
\small {24/1A Chandi Ghosh Road, Kolkata-700040, India}\\
\small{$^2$Department of Applied Mathematics, University of Calcutta}\\
\small {92 Acharya Prafulla Chandra Road, Kolkata-700009, India}}

 \date{}
\maketitle

Castorina $et$ $al.$ [1, 2] have attempted to derive logistic equations based on the formalism of phenomenological universalities, found in different field of sciences. They suggested that the logistic equations can be derived, for negative $b$, from the equation given by,\\
\begin{equation}
\frac{dy}{d\tau}=\alpha_2y^p-\beta_2y
\end{equation}
where, $\alpha_2=(1+b)/b$, $p=1-b$ and $\beta_2=1/b$. The usual logistic growth equation is obtained for $p=2$ [1, 2]. Here we point out that the conditions do not fulfill the characteristic features of logistic growth equation. The changes take place in the coefficients ($\alpha_2$ and $\beta_2$) with the change in $p$ (or $b$) is not considered here. The usual logistic growth equation is expressed generally in terms of $carrying~capacity$ that is not considered in this phenomenological description. Even logistic growth equations do not belongs to the phenomenological class described as U2 [1, 2]. Different cases obtained for different values of $p$ ((for b$<$0) are considered in the following section.\\
The condition $p=2$ follows $b=-1$, $\alpha_2=0$ and $\beta_2=-1$. The characteristic equation, derived from equation (1), of corresponding system is as follows,\\
\begin{equation}
\frac{dy}{d\tau}=y
\end{equation}
The condition $b<-1$ implies that $\alpha_2>0$, $p>2$ and $-1<\beta_2<0$. The governing equation of the corresponding system is\\
\begin{equation}
\frac{dy}{d\tau}=\alpha_2y^p+\mid\beta_2\mid y
\end{equation}
The conditions $\alpha_2<0$, $1<p<2$ and $\beta_2<-1$ are valid for $-1<b<0$. The conditions immediately follow,\\
\begin{equation}
\frac{dy}{d\tau}=\mid\beta_2\mid y-\mid \alpha_2\mid y^p
\end{equation}
Equation (2) and (3) do not represent logistic equations. Apparently it appears like that the equation (3) follows logistic growth equations, but the associated condition $a(0)=1$ is in contradiction with the characteristic feature of logictic equation, that is also true for equation (2) and (3).\\
We have repeated the calculation of the phenomenological class, as proposed by Castorina $et$ $al.$, $\varphi=b_1a+b_2a^2$ (instead of $\varphi=a+ba^2$), representing the class in a more generalized manner, with the assumption that $y(0)=1$ and $a(0)=b_1(1-\frac{1}{K})$; where $K$ is the carrying capacity. It is found that the system follows,\\
\begin{equation}
\frac{dy}{d\tau}=\alpha y^\sigma-\beta y
\end{equation}
Where, $\sigma=1-b_2$, $\alpha=b_1(1+\gamma)/b_2$, $\gamma=b_2(K-1)/K$ and $\beta=b_1/b_2$; with the solution,\\
\begin{equation}
y=[1+\gamma-\gamma exp(-b_1\tau)]^{1/b_2}
\end{equation}
\begin{figure}[htp]
\includegraphics[width=10cm, height= 5cm]{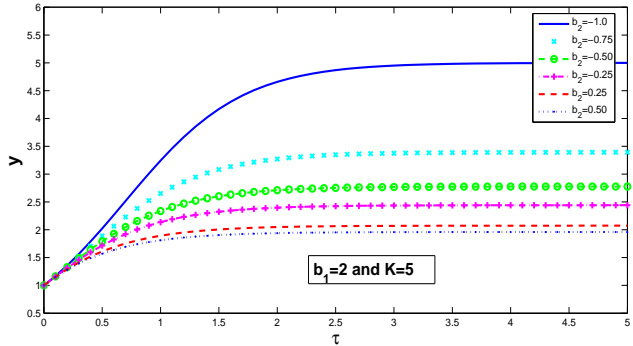}
\caption{Growth curves for $K=5$ and $b_1=2$. From the top to the bottom the values of the parameter $b_2$ are $-1.0$, $-0.75$, $-0.5$, $-0.25$, $0.25$, $0.5$. The solid curve ($b_2=-1.0$) corresponds to the usual logistic growth, while the dashed line ($b_2=0.25$) represents biological growth following West-type equation.}
\label{name}
\end{figure}
It is easy to show that equation (5) and (6) correspond to the well-known West-type equation of biological growth [3] for the condition $b_1>0$ and $b_2=0.25$. Other biological growth processes [4] may be described for $0< b_2< 0.33$. The same leads to logistic equation for the condition $b_1>0$ and $b_2<0$. The usual logistic equation is found for $b_2=-1$. When $\gamma=-1$ , the system follows exponential growth for which $b_2<0$ for $K>1$. Such behavior is not observed for $b_2>0$.\\
Therefore, the phenomenological class described above is not similar to the U2 class [1, 2]. It is because of the fact that $\varphi$ is similar in nature with U2, but the initial conditions are different from U2. In brief, we have defined a new class of phenomenological universalities that leads to logistic equation as well as West-type biological growth equation based on the values of coefficients of phenomenological description.\\

\end{document}